\documentclass[
 reprint,
 superscriptaddress,
showpacs,preprintnumbers,
 amsmath,amssymb,
 aps,
 pra
]{revtex4-1}

\usepackage{amsmath,amssymb,bbm,epsfig}
\usepackage{enumerate}

\usepackage{color}
\usepackage[usenames,dvipsnames,svgnames,table]{xcolor}
\usepackage[colorlinks=true,
            linkcolor=blue,
            urlcolor=blue,
            citecolor=blue]{hyperref}

\newcommand{\be}{\begin{equation}}
\newcommand{\ee}{\end{equation}}
\newcommand{\ba}{\begin{array}}
\newcommand{\ea}{\end{array}}
\newcommand{\bqa}{\begin{eqnarray}}
\newcommand{\eqa}{\end{eqnarray}}

\newcommand{\um}{\mathbbm{1}}

\newcommand{\vc}{\boldsymbol{c}}

\usepackage{pdfpages}

\begin{document}

\title{Tunable Chern insulator with shaken optical lattices}

\author{Albert Verdeny}
\author{Florian Mintert}

\affiliation{Department of Physics, Imperial College London, London SW7 2AZ, United Kingdom}
\affiliation{Freiburg Institute for Advanced Studies, Albert-Ludwigs-Universit\"at, Albertstrasse 19, 79104 Freiburg, Germany}

\begin{abstract}
Driven optical lattices permit the engineering of effective dynamics with well-controllable tunneling properties.
We describe the realization of a tunable a Chern insulator by driving particles on a shaken hexagonal lattice with optimally designed polychromatic driving forces. 
Its implementation does not require shallow lattices, which favors the study of strongly-correlated phases with non-trivial topology.
\end{abstract}

\maketitle

\hypersetup{%
    pdfborder = {0 0 0}
}

The manipulation of quantum systems has reached levels of accuracy that allow controlled variation of their properties.
This permits extremely precise investigations of physical processes and opens up the opportunity to design materials for scientific and economic applications.
In particular, the last years have witnessed an increasing interest in topological phases of matter largely motivated by promising applications such as topological quantum computing \cite{Nayak08} or physical phenomena such as the quantum spin Hall effect \cite{Sinova04,Kane05,Kane05Z2}.

Systems with topological properties can be classified with topological invariants whose non-zero values indicate non-trivial phases \cite{Hasan10}. Pioneer work in this field was done in the context of the integer quantum Hall effect, when it was shown that the quantization of the Hall conductance is directly related to the first Chern invariant \cite{Thouless82}. Thereafter, Haldane demonstrated with a model of particles tunneling on an hexagonal lattice \cite{Haldane88} that a net magnetic field is not necessary for the quantization of the Hall conductance, owing to the topological nature of the model. Remarkably, the Haldane model has been recently experimentally implemented with shaken optical lattices \cite{Jotzu14}. Nonetheless, the exploration of its topological diagram crucially requires shallow lattices since preexisting next-nearest-neighbor tunneling in the undriven system is necessary.

Periodically driven lattices offer an extraordinary platform to engineer controlled dynamics.
A number of theoretical \cite{Jaksch03,Eckardt05,Inoue10,Hauke12,Delplace13,Baur14,Bukov14ar} and experimental \cite{Lignier07,Lin09,Aidelsburger11,Struck11,Jotzu14} works have demonstrated the possibility to modify the system dynamics in a controlled fashion and to generate diverse effects that include coherent destruction of tunneling \cite{Lignier07} and the creation of synthetic magnetic fields \cite{Lin09,Aidelsburger11,Struck11} or topological properties \cite{Jotzu14}.

Even though the effective dynamics that a driven system undergoes crucially depend on the specific time-dependent driving force, rather simple driving forces are usually employed. Nevertheless, as demonstrated in various fields, including chemistry \cite{Rabitz00,Koch12,Hoff12}, nuclear magnetic resonance \cite{Conolly86,Khaneja05}, quantum information \cite{Timoney08,Platzer10}, and many-body systems \cite{Doria11}, essentially any desired dynamics can be induced with the appropriate choice of polychromatic driving at desired instances of time or during an extended time-window \cite{Bartels13,Verdeny14}.

In this Letter we show an optimal implementation of a Chern insulator by driving non-interacting particles on an hexagonal lattice with polychromatic driving. We demonstrate how the entire topological diagram can be accessed with deep optical lattices and suitably chosen driving forces.

Driven systems can be used as quantum simulators due to the possibility to approximate their dynamics in terms of a time-independent Hamiltonian. 
According to Floquet theorem \cite{Floquet83}, the time-evolution operator of a periodic Hamiltonian $H(t)=H(t+T)$ can be written as 
\begin{eqnarray}\label{decomp}
U(t)=U_F^\dagger(t)e^{-iH_{\rm eff}t}U_F(0),
\end{eqnarray}
where $U_F(t)$ is a $T$-periodic unitary and $H_{\rm eff}$ defines a time-independent effective Hamiltonian.
The distance between the exact dynamics induced by $H(t)$ and the approximate dynamics $U_{\rm eff}(t)=e^{-iH_{\rm eff}t}$ induced by $H_{\rm eff}$ is bounded $||U(t)-U_{\rm eff}(t)||\leq ||\um-U_F^\dagger(t)||+||\um-U_F(0)||$.
Consequently, if the unitary $U_F(t)$ is sufficiently close to the identity during an entire period, the dynamics of the system can be very well captured by $U_{\rm eff}$ for all times.
This condition is typically satisfied in a suitable fast-driving regime, where the driving frequency $\omega=2\pi/T$ is the largest energy scale of the system. 
The effective Hamiltonian can then be found in a perturbative expansion $H_{\rm eff}=H_{\rm eff}^{(0)}+H_{\rm eff}^{(1)}+\dots$ in powers of $\omega^{-1}$ using different methods \cite{Magnus54,Schirley65,Rahav03,Verdeny13}.
The lowest-order term $H_{\rm eff}^{(0)}$ corresponds to the average or static Fourier component of the Hamiltonian. Higher-order terms, on the other hand, depend on the particular choice of gauge $U_F(0)$. For convenience, we choose the gauge \cite{Goldman14} that leads to 
\begin{eqnarray} \label{heff1}
H_{\rm eff}^{(1)}&=& \dfrac{1}{\omega}\sum_{n=1}^\infty \dfrac{1}{n} [H_n,H_{-n}],
\end{eqnarray}
with the Fourier components $H_n=\frac{1}{T}\int_0^T H(t) e^{-in\omega t} dt$.

We consider spinless, non-interacting particles on a shaken hexagonal lattice described by the   Hamiltonian
\begin{eqnarray}\label{drivenhexagonal}
H(t)&=&\sum_{i} \Big( \vc^\dagger_{\textbf{r}_i}  d_0(t)\ \vc_{\textbf{r}_i}+\sum_{k=1}^2 (\vc^\dagger_{\textbf{r}_i+\textbf{b}_k}  d_k(t)\ \vc_{\textbf{r}_i}+\mbox{H.c.})\Big),\nonumber \\
\end{eqnarray}
with the vector creation and annihilation operators $\vc^\dagger_{\textbf{r}_i}=(c^\dagger_{A,i}, c^\dagger_{B,i})$ and $\vc_{\textbf{r}_i}=(c_{A,i}, c_{B,i})^T$ satisfying the usual (anti)commutator relations, and the time-dependent matrices
\begin{eqnarray}
d_0(t)&=&\left( \begin{matrix}
\Delta&g_{\textbf{a}_3}^*(t)\\
 g_{\textbf{a}_3}(t)&-\Delta
\end{matrix}\right), \\ 
d_1(t)&=&\left( \begin{matrix}
0&0\\
g_{\textbf{a}_2}(t)&0
\end{matrix}\right), \ d_2(t)=\left( \begin{matrix}
0&g_{\textbf{a}_1}^*(t)\\
0&0
\end{matrix}\right)
\end{eqnarray}
with energy offset $\Delta$.
The summation in Eq. (\ref{drivenhexagonal}) is performed over the positions  $\textbf{r}_i$ of all unit cells of the hexagonal lattice, which we consider to be infinite or with periodic boundary conditions. 
The vectors $\textbf{b}_1=a(\sqrt{3},0)$ and $\textbf{b}_2=\frac{a}{2}(-\sqrt{3},3)$ correspond to two primitive vectors that span the underlying triangular Bravais lattice, with the distance $a$ between nearest-neighbor (NN) sites.
The time-dependent rates that characterize the NN tunneling are given by $g_{\textbf{a}_k}(t)=j_0 \ e^{ i \chi_k(t)}$, with $\chi_k(t)=\int_0^t d\tau \ \textbf{F}(\tau)\cdot\textbf{a}_k -\frac{1}{T}\int_0^T dt \int_0^t d\tau\ \textbf{F}(\tau)\cdot\textbf{a}_k $ in terms of the driving force $\textbf{F}(t)$ and the real tunneling amplitudes $j_{0}$ of the undriven system. In general, the time-dependent tunneling rates depend on the direction of tunneling defined through the vectors that connect neighboring sites $\textbf{a}_1=\frac{a}{2}(\sqrt{3},1)$, $\textbf{a}_2=\frac{a}{2}(-\sqrt{3},1)$ and $\textbf{a}_3=-\textbf{a}_1-\textbf{a}_2$.

As theoretically expected and experimentally confirmed \cite{Jotzu14}, the dynamics of the system in a fast driving regime can be captured very well by the truncated effective Hamiltonian
\begin{eqnarray}\label{Hdh}
H_{\rm dh}=H_{\rm eff}^{(0)}+H_{\rm eff}^{(1)}.
\end{eqnarray}
Since the leading-order effective Hamiltonian is given by the average of $H(t)$ in Eq. (\ref{drivenhexagonal}), $H_{\rm eff}^{(0)}$ contains the same tunneling processes as the undriven system: the on-site energies $\pm \Delta$ remain invariant and the effective tunneling rates become the directionality-dependent quantities $g_{\textbf{a}_k}^0=\frac{1}{T}\int_0^T dt\ g_{\textbf{a}_k}(t)$.

The first-order term $H_{\rm eff}^{(1)}$ given by Eq. (\ref{heff1}) reads
\begin{eqnarray}
H_{\rm eff}^{(1)}&=&\sum_{i} \sum_{k=0}^3\vc^\dagger_{\textbf{r}_i+\textbf{b}_k} h^{\rm eff}_{k} \ \vc_{\textbf{r}_i}+\mbox{H.c.},
\end{eqnarray}
where $\textbf{b}_0=0$, $\textbf{b}_3=-\textbf{b}_1-\textbf{b}_2$. The effective matrices $h^{\rm eff}_{k}=\mbox{diag}(\tau_k,-\tau_k)$ with
$\tau_0=\sum_{i=1}^3w(\textbf{a}_i,-\textbf{a}_i)$,
$\tau_1=w(\textbf{a}_2,-\textbf{a}_3)$
$\tau_2=w(\textbf{a}_3,-\textbf{a}_1)$ and
$\tau_3=w(\textbf{a}_1,-\textbf{a}_2)$
are defined in terms of 
\begin{eqnarray}\label{kappa1}
w(\textbf{a}_i,\textbf{a}_j)&=& \sum_{n=1}^\infty \dfrac{1}{n\omega}\left(g_{\textbf{a}_i}^{-n} g_{\textbf{a}_j}^{n}-g_{\textbf{a}_j}^{-n}g_{\textbf{a}_i}^n\right),
\end{eqnarray}
with the Fourier components $g_{\textbf{a}_j}^n=\frac{1}{T}\int_0^T g_{\textbf{a}_j}(t)e^{-in\omega}$.
The rates $\tau_k$, $k=1,2,3$, describe effective next-nearest-neighbor (NNN) tunneling processes that result from a virtual tunneling process over a neighboring site. 
The relative sign in $h^{\rm eff}_{k}$ between the different rates $\tau_k$ and $-\tau_k$ is a fundamental symmetry that is independent of the specific driving force $\textbf{F}(t)$.  

Due to this symmetry, the emergent NNN tunneling rates discussed above are, in general, not equivalent to those of the Haldane model \cite{Haldane88}, where the two tunneling rates are complex conjugated with respect to each other. Only for purely imaginary rates $\tau_k^*=-\tau_k$, $k=1,2,3$, do the NNN rates of the two models coincide.
Consequently, it is fundamentally impossible to implement the full topological diagram of the Haldane model via lattice shaking without non-vanishing real NNN tunneling rates in the undriven system. 

Despite the differences between the Haldane model Hamiltonian and the effective Hamiltonian in Eq. (\ref{Hdh}), the two models share similar topological properties.
As we shall later demonstrate, it is possible to find a driving force yielding isotropic tunneling rates, namely $g_{\textbf{a}_k}^0=j_1$ and $\tau_k=j_2 e^{i\phi}$ for all directions $k=1,2,3$, where $j_1$ and $j_2$ are positive real numbers and $\phi$ is defined in the interval $(-\pi,\pi]$.
The effective Hamiltonian can then be written in quasimomentum space as $H_{\rm dh}=\sum_\textbf{k} \vc_\textbf{k}^\dagger H(\textbf{k})\vc_\textbf{k}$,
where $\vc_\textbf{k}^{(\dagger)}$ are the vector momentum creation and annihilation operators and
\begin{eqnarray}\label{hamis}
H(\textbf{k})&=& \sum_{i=1}^3 h_i(\textbf{k})\sigma_i
\end{eqnarray}
is defined in terms of the Pauli matrices $\sigma_i$ and 
\begin{eqnarray}
h_1(\textbf{k})&=&j_1 \left( 1+\cos(\textbf{k}\cdot \textbf{b}_1)+\cos(\textbf{k}\cdot \textbf{b}_2) \right), \\
h_2(\textbf{k})&=&j_1 \left(\sin(\textbf{k}\cdot \textbf{b}_1)-\sin(\textbf{k}\cdot \textbf{b}_2)  \right),\\
h_3(\textbf{k})&=&\Delta+2j_2 \sum_{i=1}^3 \cos(\textbf{k}\cdot \textbf{b}_i+\phi) .
\end{eqnarray}
\begin{figure}[t]
  \centering
    \includegraphics[width=0.48\textwidth]{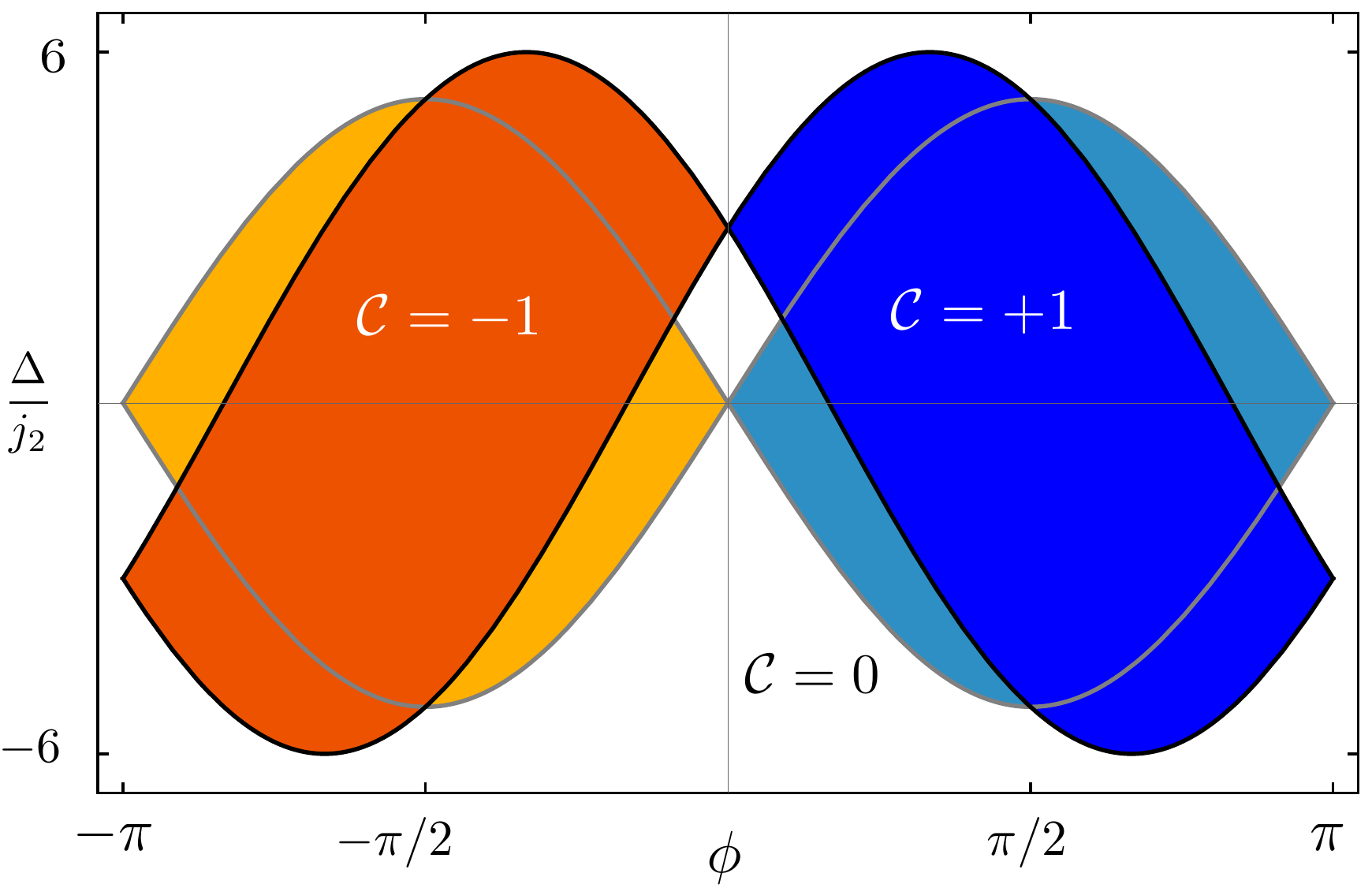} 
      \caption{Phase diagram of the isotropic effective Hamiltonian in Eq. (\ref{hamis}) (black lines and dark colors) overlapped with the phase diagram of the Haldane model \cite{Note1} (gray lines and lighter colors), giving the Chern number $\mathcal{C}$ of the lowest ernergy band as a function of the phase $\phi$ and ratio $\Delta/j_2$. Orange represents $\mathcal{C}=-1$, blue $\mathcal{C}=1$ and white $\mathcal{C}=0$. For $\phi=\pm \pi/2$ the Chern number of the two Hamiltonians coincide independently of $\Delta/j_2$.}\label{diagram}
\end{figure}
The topological diagram of this model, displaying the Chern number \cite{Hasan10} as a functions of the Hamiltonian parameters $\Delta/j_2$ and $\phi$, can be readily calculated \cite{Sticlet12} and it is shown in Fig. \ref{diagram}.
The transition between different topological phases -- indicated with a solid black line -- corresponds to parameters of the Hamiltonian for which the gap between the two energy bands $\epsilon_{\pm}(\textbf{k})=\sqrt{h_1^2+h_2^2+h_3^2}$ closes.
For comparison, we also display the analogous topological diagram of the isotropic Haldane model Hamiltonian \footnote{\label{note1}Defined in terms of the Hamiltonian $H_{\rm H}(\textbf{k})=h'_0(\textbf{k})\um+\sum_{i=1}^3 h'_i(\textbf{k})\sigma_i$ with $h'_0(\textbf{k})=2j_2 \cos(\phi)\sum_{i=1}^3\cos(\textbf{k}\cdot\textbf{b}_i)$, $h'_{1,2}(\textbf{k})=h_{1,2}(\textbf{k})$ and $h'_{3}(\textbf{k})=h_{3}(\textbf{k})-h'_{0}(\textbf{k})$.}. Only for $\phi=\pm\pi/2$ do the two Hamiltonians coincide, consistently with the diagram in Fig. \ref{diagram}. 

In order to assess to what extend the entire parameter regime of the topological diagram can be realistically explored, it is necessary to correctly identify driving forces $\textbf{F}(t)$ that lead to isotropic effective tunneling rates with controllable amplitudes and phase $\phi$.
Since the topological energy bands emerge as a consequence of the interplay between the NN and NNN tunneling processes, it is important that the relative effect of the NNN tunneling with respect to NN tunneling, given by the ratio $j_2/j_1$, is sufficiently large. Nevertheless, these two tunneling processes are of different orders of magnitude, since $j_1\sim j_0$ and $j_2\sim j_0^2/\omega$. As the ratio $j_2/j_1$ is proportional to $j_0/\omega$, it could easily be increased through a decrease of the driving frequency $\omega$. This, however, could compromise the validity of the high frequency expansion of the effective Hamiltonian.
For this reason, we consider a small fixed ratio $j_0/\omega$ to be determined according to the experimental setup and aim at finding a driving force with a set of parameters $\textbf{p}$ that maximize the proportionality factor $\frac{j_2}{j_1} \frac{\omega}{j_0}$ between $j_2/j_1$ and $j_0/\omega$. 
Since the amplitude $j_2$ is directly related to  $\phi$, we introduce $\phi=\phi_{\rm tg}$ as a constraint for the maximization, where $\phi_{\rm tg}$ is the desired phase that we target.
Additionally, $j_1$ should be sufficiently large with respect to the bare tunneling rate $j_0$ in order to avoid that the effective tunneling processes appear at the expense of slowing down the dynamics as compared to the undriven system.
We therefore introduce the additional constraint $j_1/j_0\geq r_{\rm th}$, where the threshold value $r_{\rm th}$ can be chosen from the interval $0\le r_{\rm th}\le 1$.

We thus aim at finding a driving force targeting:
\begin{enumerate}
\item[(i)] Isotropy $g_{\textbf{a}_k}^0=j_1$ and $\tau_k=j_2e^{i\phi}$, $k=1,2,3$.
\item[(ii)] Controlability of the phase $\phi$.
\item[(iii)] Enhancement of the NNN tunneling rates through the maximization
 \begin{eqnarray}\label{R}
R(\phi_{\rm tg},r_{\rm th})=\left\{\max_\textbf{p}\dfrac{j_2}{j_1} \frac{\omega}{j_0} \ \Big|\ \dfrac{j_1}{j_0}\geq r_{\rm th} \mbox{ ; } \phi=\phi_{\rm tg}\right\}
\end{eqnarray}
performed over a set of free driving parameters $\textbf{p}$.
\end{enumerate}

For a monochromatic driving force \cite{Jotzu14,Oka09,Kitagawa11,Grushin14}, the effective NNN tunneling rates become purely imaginary and, thus, only the points $\phi=\pm \pi/2$ in Fig. \ref{diagram} can be accessed. This strong limitation can be overcome by specifically designing the driving pulse to satisfy our requirements. 
Despite the highly non-linear dependence that the tunneling rates have on the driving force, we find analytic expressions for $g_{\textbf{a}_k}^0$ and $\tau_k$ in terms of driving parameters by using $N$-dimensional Bessel functions \footnote{See Supplemental Material for an details on multidimensional Bessel functions, derivation of the isotropic tunneling rates and optimal results for a force with three Fourier components.}.

This allows us to identify the structure of two driving forces $\textbf{F}_+(t)$ and $\textbf{F}_-(t)$ that lead to isotropic tunneling rates independently of their free parameters \cite{Note2}.  
Consequently, the requirement (i) above is automatically satisfied \footnote{In fact, the forces $\textbf{F}_\pm(t)$ in Eq. (\ref{forcepm}) can lead to complex nearest-neighbor tunneling $j_1$, but it is possible to perform a local unitary transformation that eliminates the complex phase without affecting the next-nearest-neighbor tunneling rates. In other words, the complex phase of $j_1$ can be trivially incorporated in the definitions of the creation and annihilation operators.}, which allows the examination of the subsequent target properties.
The general form of the driving forces $\textbf{F}_\pm(t)$ containing $N$ different frequency harmonics reads

\begin{eqnarray}\label{forcepm}
\textbf{F}_\pm(t) &=&\sum_{n=1}^N A_n \left(\cos(\omega_n t-\delta_n)\textbf{e}_1 +\cos(\omega_n t-\delta^{\pm}_n)\textbf{e}_2\right)\nonumber\\
\end{eqnarray}
with the two perpendicular vectors $\textbf{e}_1=(\textbf{a}_1-\textbf{a}_2)/\sqrt{3}$ and $\textbf{e}_2=-\textbf{a}_3$, the relative phases
$\delta^{\pm}_n =\delta_n\pm(-1)^n\pi/2$ and the frequencies
$\omega_m=\frac{1}{4} \left(6m-(-1)^{m}-3\right)\omega$,
which parametrize all positive integer multiples of $\omega$ except those that are multiples of $3\omega$. 
Because the overall phase of the driving force is irrelevant in the fast-driving regime, we choose $\delta_1=0$ in the following. The remaining $2N-1$ driving parameters comprise the set $\textbf{p}$ and need to be chosen so that the requirements (ii) and (iii) are satisfied.

\begin{figure}[t]
\includegraphics[width=0.45\textwidth]{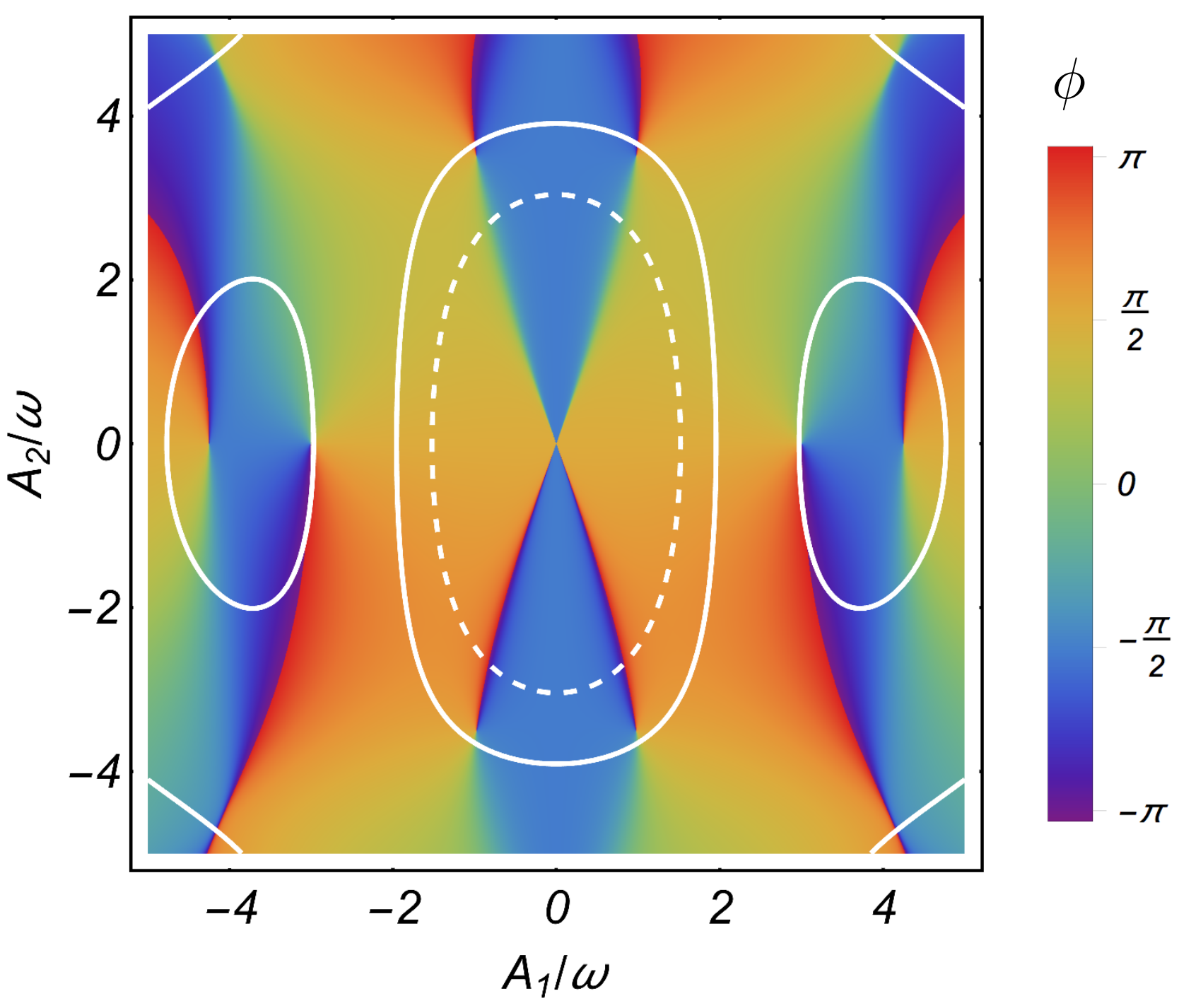} 

\caption{Phase $\phi$ of the complex effective next-nearest-neighbor tunneling rates as a function of $A_1/\omega$ and $A_2/\omega$ for a force $\textbf{F}_+(t)$ with $N=2$ and $\delta_2=\pi/2$. All phases can be explored with a suitable choice of $A_1$ and $A_2$. 
The dashed and solid white lines indicate the contour lines for $j_1/j_0=0.5$ and $j_1/j_0=0.25$ and limit the region of accessible phases given a constraint with $r_{\rm th}=0.5$ or $r_{\rm th}=0.25$, respectively.
}\label{phase}
\end{figure} 

In order to ease an experimental implementation, we consider the simplest polychromatic force with $N=2$, which contains three driving parameters: $A_1$, $A_2$ and $\delta_2$. We analytically find that if $\delta_2=0,\pm \pi$, the real part of $\tau_k$ again vanishes, similarly to the monochromatic case. However, for $\delta_2\neq0,\pm \pi$ and appropriate choice of driving amplitudes the entire range of phases $\phi\in (-\pi,\pi]$ can be realized, satisfying thus the requirement (ii), as illustrated in Fig. \ref{phase}. 
\begin{figure}[t]
  \centering
    \includegraphics[width=0.45\textwidth]{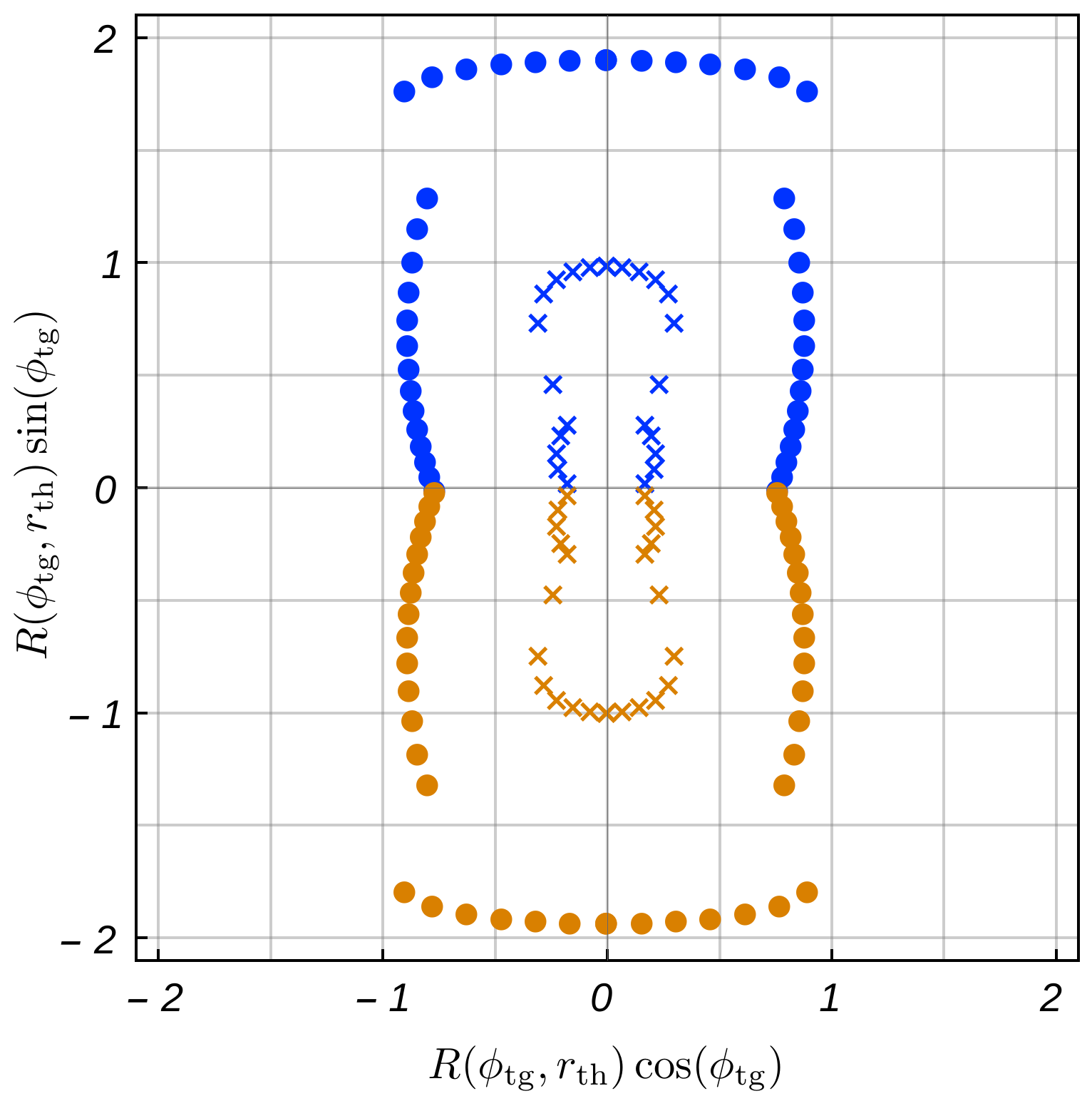}
      \caption{Plot of the real and imaginary part of $R(\phi_{\rm tg},r_{\rm th})e^{i\phi_{\rm tg}}$ as a function of a discrete set of target phases $\phi_{\rm tg}$ and for two different values of $r_{\rm th}$. In polar coordinates, the radius and argument of each data point coincide with $R(\phi_{\rm tg},r_{\rm th})$ and $\phi_{\rm tg}$, respectively. Dots correspond to $r_{\rm th}=0.25$, and crosses to $r_{\rm th}=0.5$. The results in blue have been obtained with a driving force $\textbf{F}_+(t)$ and the results in orange with $\textbf{F}_-(t)$.}\label{CP}
\end{figure}

The maximization in the point (iii) leads to a significant enhancement of the effective NNN tunneling for any desired phase, but considerably different results are obtained depending on the phase $\phi_{\rm tg}$ and threshold rate $r_{\rm th}$ that we target.  
In order to discuss this behavior, we plot in Fig. \ref{CP} the real and imaginary part of $R(\phi_{\rm tg},r_{\rm th})e^{i\phi_{\rm tg}}$ for two different values of $r_{\rm th}$ and for a discrete set of angles $\phi_{\rm tg}\in (-\pi,\pi]$. Each data point is obtained with a numerical constrained optimization over the set of free parameters $\textbf{p}=(A_1/\omega, A_2/\omega, \delta_2)$. Overall, we observe two main features.

{\em First},  for a given threshold value $r_{\rm th}$, the largest values of $R(\phi_{\rm tg},r_{\rm th})$ are obtained for a range of phases close to $\pm \pi/2$. The lowest values correspond to phases $0$ and $\pm\pi$, for which effective Hamiltonian $H_{\rm dh}$ is time-reversal invariant. This indicates that experimentally it is easier to access the areas of the topological diagram in Eq. (\ref{diagram}) that are near $\phi=\pm\pi/2$.
Noteworthy, we find that for  $\phi_{\rm tg}=\pm \pi/2$ the optimal solution of the two-frequency pulse reduces to a monochromatic force, i.e. $A_2=0$, independently of $r_{\rm th}$.
Nonetheless, the maximum of $R(\phi_{\rm tg},r_{\rm th})$ does not always correspond to $\phi_{\rm tg}=\pm\pi/2$ for a fixed $r_{\rm th}$, as can be seen in the results for $r_{\rm th}=0.25$ in Fig. \ref{CP}.

{\em Second}, the lower the threshold value $r_{\rm th}$ is for a fixed target phase $\phi_{\rm tg}$, the larger $R(\phi_{\rm tg},r_{\rm th})$ can be, as a larger region in the parameter space, given by the driving parameters, can be accessed (see contour lines in Fig. \ref{phase}). 
Thus, there is a trade-off between lower values of the threshold $r_{\rm th}$ for the ratio $j_1/j_0$ and higher relative enhancement $\frac{j_2}{j_1} \frac{\omega}{j_0}$ of NNN tunneling.
It is thus advisable to choose a small value of $r_{\rm th}$, provided that it is sufficiently large so that tunneling is dominant in the dynamics of the system and processes like interaction or heating can be neglected on the time-scale on which tunneling occurs.

As a result of the maximization in Eq. (\ref{R}), the set of optimal driving parameters that yield the maximum enhancement of the NNN tunneling are also obtained. 
We find that for all the results shown in Fig. \ref{CP}, the corresponding driving amplitudes are of a similar order of magnitude as the driving frequency, specifically $|A_i/\omega|<3.5$, $i=1,2$. Additionally, we observe a discontinuous behavior of the optimal driving parameters as a function of $\phi_{\rm tg}$, which leads to a discontinuity in $R(\phi_{\rm tg},r_{\rm th})$, see $r_{\rm th}=0.25$ results in Fig. \ref{CP}. This can be understood in terms of the constraint $j_1/j_0>r_{\rm th}$, which restricts the parameter space to disjoint regions, as can be seen in the contour line $j_1/j_0=0.25$ in Fig. \ref{phase}. Depending on the targeted phase, the driving parameters might change from one region to another, yielding a discontinuity in the driving parameters and in the corresponding value of $R(\phi_{\rm tg},r_{\rm th})$.

Increasing the number of frequency harmonics of the force $\textbf{F}_{\pm}(t)$, more driving parameters become accessible. These additional degrees of freedom can be exploited e.g. to further increase the maximum value of $R(\phi_{\rm tg},r_{\rm th})$ targeting specific phases, as we demonstrate for a driving force with $N=3$ \cite{Note2}.

Our results show that even a very low number of frequency components is sufficient to completely outperform the monochromatic driving and yield significant NNN tunneling rates for any phase $\phi$.
This exemplifies the fact that the usually-considered monochromatic driving can strongly limit the accessible effective dynamics and that only suitably chosen driving protocols can enable the exploration of the entire accessible dynamics.
Finally, we believe that the possibility to experimentally implement the model in Eq. (\ref{hamis}) in deep optical lattices naturally warrants further theoretical investigations, e.g. on the emergence of fractional quantum Hall effect \cite{Neupert12} when interactions are taken into account.

We are indebted to Julian Struck, Juliette Simonet, Gregor Jotzu, Andreas Mielke and Elizabeth von Hauff for stimulating discussions or feedback on the manuscript.
Financial support by the European Research Council within the project ODYCQUENT is gratefully acknowledged.

\bibliography{mybib}

\begin{widetext}
\clearpage


\section*{Supplementary material}

# Supplemental Material for *Tunable Chern insulator with shaken optical lattices*

Albert Verdeny[1,2] and Florian Mintert[1,2]

[1]*Department of Physics, Imperial College London, London SW7 2AZ, United Kingdom*
[2]*Freiburg Institute for Advanced Studies, Albert-Ludwigs-Universität, Albertstrasse 19, 79104 Freiburg, Germany*

This Supplemental Material is divided into three sections. In Sec. I, we introduce the mathematical tools necessary to derive analytical expressions for the tunneling rates of the effective Hamiltonian of shaken lattices. In Sec. II, we demonstrate that the driving forces introduced in Eq. (14) lead to isotropic tunneling rates. Finally, in Sec. III we provide the optimal results corresponding to a force with three Fourier components.

## I. MULTIDIMENSIONAL BESSEL FUNCTIONS

The effective tunneling rates of the Hamiltonian $H_{\rm dh}$ in Eq. (6) are given in terms of the Fourier components $g_{\mathbf{a}_i}^n$, which are defined through

$$j_0 e^{i\chi_k(t)} = \sum_{n=-\infty}^{\infty} g_{\mathbf{a}_i}^n e^{in\omega t} \quad (S1)$$

with a $T$-periodic function $\chi_k(t)$, as described in the main text. Motivated by the need of analytical expressions to target the properties described in the points (i)-(iii), we introduce multidimensional Bessel functions as a generalization of ordinary and two-dimensional Bessel functions [1].

Ordinary $n$-th order Bessel functions $\mathcal{J}_n(a)$ can be defined as Fourier components of the generating function

$$e^{ia\sin(b)} = \sum_{n=-\infty}^{\infty} \mathcal{J}_n(a) e^{inb}. \quad (S2)$$

Here, we generalize the Bessel functions introduced above and consider $N$-dimensional Bessel functions $\mathcal{J}_n^{\mathbf{z}}(\mathbf{c};\mathbf{d})$ defined by the generating function

$$e^{i\sum_{k=1}^N c_k \sin(z_k\tau-d_k)} = \sum_{n=-\infty}^{\infty} \mathcal{J}_n^{\mathbf{z}}(\mathbf{c};\mathbf{d}) e^{in\tau}, \quad (S3)$$

with the real vectors $\mathbf{c} = (c_1,\ldots,c_N)$ and $\mathbf{d} = (d_1,\ldots,d_N)$, and the integer vector $\mathbf{z} = (z_1,\ldots,z_N)$. The Fourier transform of Eq. (S3) leads to the integral representation

$$\mathcal{J}_n^{\mathbf{z}}(\mathbf{c};\mathbf{d}) = \frac{1}{2\pi} \int_0^{2\pi} d\tau e^{i\sum_{k=1}^N c_k \sin(z_k\tau-d_k)-in\tau}. \quad (S4)$$

Inserting Eq. (S2) into Eq. (S4) permits to expand $N$-dimensional Bessel functions in terms of ordinary one-dimensional Bessel functions

$$\mathcal{J}_n^{\mathbf{z}}(\mathbf{c};\mathbf{d}) = \sum_{\mathbf{x}} \mathcal{J}_{x_1}(c_1) \cdot \ldots \cdot \mathcal{J}_{x_N}(c_N) e^{-i\mathbf{d}\cdot\mathbf{x}}$$
$$\frac{1}{2\pi} \int_0^{2\pi} d\tau e^{i\tau(\mathbf{z}\cdot\mathbf{x}-n)} \quad (S5)$$
$$= \sum_{\mathbf{x}} \mathcal{J}_{x_1}(c_1) \cdot \ldots \cdot \mathcal{J}_{x_N}(c_N) e^{-i\mathbf{d}\cdot\mathbf{x}} \delta_{\mathbf{z}\cdot\mathbf{x},n} \quad (S6)$$
$$= \sum_{\mathbf{s}} \mathcal{J}_{s_1}(c_1) \cdot \ldots \cdot \mathcal{J}_{s_N}(c_N) e^{-i\mathbf{d}\cdot\mathbf{s}}, \quad (S7)$$

where $\mathbf{x} = (x_1,\ldots,x_N)$ denotes all possible integer vectors and $\mathbf{s} = (s_1,\ldots,s_N)$ all integer solutions of the Diophantine (i.e. integer) equation

$$\mathbf{z}\cdot\mathbf{s} = z_1 s_1 + \cdots + z_N s_N = n. \quad (S8)$$

A linear Diophantine equation of the form in Eq. (S8) can be solved if and only if the greatest common divisor $\gcd(z_1,\ldots,z_N)$ divides $n$ [2]. Since the driving force will always contain the fundamental driving frequency, at least one of the elements of $\mathbf{z}$ equals 1, which implies that Eq. (S8) can be trivially solved for all $n$. Without loss of generality we order the elements such that $z_1 = 1$. The solution of Eq. (S8) then reads

$$s_1 = n - \sum_{i=2}^N z_i s_i \quad (S9)$$

where $s_2,\ldots,s_N$ are $N-1$ integer free parameters that characterize the general solution. Consequently, since $z_1 = 1$, $N$-dimensional Bessel functions can be expressed as

$$\mathcal{J}_n^{\mathbf{z}}(\mathbf{c};\mathbf{d}) = \sum_{s_2,\ldots,s_N=-\infty}^{\infty} \mathcal{J}_{n-\sum_{j=2}^N z_j s_j}(c_1) \mathcal{J}_{s_2}(c_2) \cdot \ldots \cdot \mathcal{J}_{s_N}(c_N) e^{-i\mathbf{d}\cdot\mathbf{s}}. \quad (S10)$$

A numerical evaluation of $N$-dimensional Bessel functions can be implemented by truncating the infinite sums

in Eq. (S10). Specifically, in the numeral implementation of the maximization described in the main article we examine amplitudes with magnitudes $|c_i| < 5$, for which a truncation of $|s_i|$, $i = 1, \cdots, N$, up to $\sim 7$ is sufficient. We find that, over a wide range of parameters, this implementation is computationally more efficient than other numerical approaches such as a Taylor series truncation of the generating function in Eq. (S3). Nonetheless, the dimension of the Bessel function rapidly becomes a limiting factor since the number of terms to evaluate grows with $N$. For this reason, it is convenient to work with the lowest dimension possible, which is given by the number of elements in $\mathbf{z}$ that are different. If a subset $\mathbf{z}'$ of $\mathbf{z}$ contains the same elements, i.e. $\mathbf{z}' = (z, \ldots, z)$, then the polar parametrization

$$re^{-i\alpha} \equiv \sum_i c_i' e^{-id_i'} \tag{S11}$$

allows to express

$$\sum_i c_i' \sin(z\tau - d_i') = r \sin(z\tau - \alpha), \tag{S12}$$

where $r \geq 0$ and $c_i'$, $d_i'$ are the elements of $\mathbf{c}$ and $\mathbf{d}$ associated to $\mathbf{z}'$. In this manner, the dimension of $\mathbf{z}$ can be reduced by $\dim(\mathbf{z}') - 1$. This process can be repeated until all the elements of $\mathbf{z}$ are different.

## II. ISOTROPIC TUNNELING RATES

In this section we will demonstrate how the driving forces $\mathbf{F}_\pm(t)$ introduced in Eq. (14) lead to isotropic tunneling rates $g^0_{\mathbf{a}_k}$ and $\tau_k$ defined after Eq. (6) and (7), respectively. For this purpose, we shall first find analytical expressions in terms of multidimensional Bessel functions for the Fourier components $g^n_{\mathbf{a}_k}$ of the time-dependent tunneling rates in Eq. (8).

We start by considering the force $\mathbf{F}_+(t)$. The quantity $\chi_k(t)$ introduced after Eq. (5) then reads

$$\chi_k(t) = \sum_{m=1}^N \frac{A_m}{\omega_m}(\sin(\omega_m t - \delta_m)\mathbf{a}_k \cdot \mathbf{e}_1 \\ + \sin(\omega_m t - \delta_m^+)\mathbf{a}_k \cdot \mathbf{e}_2). \tag{S13}$$

Following the same arguments as in the previous section, we introduce the polar representation

$$c_m e^{-id_m^k} = \frac{A_m}{\omega_m}(e^{-i\delta_m}\mathbf{a}_k \cdot \mathbf{e}_1 + e^{-i\delta_m^+}\mathbf{a}_k \cdot \mathbf{e}_2) \tag{S14}$$

$$= \frac{A_m}{\omega_m} e^{-i\delta_m}(\mathbf{a}_k \cdot \mathbf{e}_1 - i(-1)^m \mathbf{a}_k \cdot \mathbf{e}_2). \tag{S15}$$

Importantly, since the vectors $\mathbf{a}_1$, $\mathbf{a}_2$ and $\mathbf{a}_3$ have the same amplitude and because the chosen relative phases $\delta_m^{(\pm)}$ satisfy $\delta_m^+ - \delta_m = (-1)^m \pi/2$, the quantities $c_m$ do not depend on the direction specified by the index $k$.

Moreover, from Eq. (S15) we find that the directionality-dependent phases $d_m^k$ can be expressed as

$$d_m^k = \delta_m + (-1)^m \phi_k, \tag{S16}$$

where $\phi_k$ denotes the angle of the vector $\mathbf{a}_k$ with respect to $\mathbf{e}_1$ and they read $\phi_1 = \pi/6$, $\phi_2 = 5\pi/6$ and $\phi_3 = 3\pi/2$ with the convention introduced in the main text after Eq. (5).

Using this representation, the number of terms in Eq. (S13) is reduced from $2N$ to $N$ such that

$$\chi_k(t) = \sum_{m=1}^N c_m \sin(\omega_m t - d_m^k). \tag{S17}$$

With isotropy in the undriven system, i.e. $j_{\mathbf{a}_k} = j_0$, $k = 1, 2, 3$, the Fourier components $g^n_{\mathbf{a}_k}$ can be expressed in terms of $N$-dimensional Bessel functions as

$$g^n_{\mathbf{a}_k} = j_0 \; \mathcal{J}_n^{\mathbf{z}}(\mathbf{c}; \mathbf{d}^k), \tag{S18}$$

where $\mathbf{c} = (c_1 \cdots, c_N)$, $\mathbf{d}^k = (d_1^k \cdots, d_N^k)$ and $\mathbf{z} = (z_1, \cdots, z_N)$ of the Bessel functions, with $z_m = \omega_m/\omega$.

The directionality dependence of the Fourier components $g^n_{\mathbf{a}_k}$ appears only in the exponential $e^{-i\mathbf{d}^k \cdot \mathbf{s}}$ of the Bessel function, see Eq. (S10). In fact, due to Eq. (S16), the only possible directionality dependence originates from $\mathbf{p}^k \equiv (-\phi_k, \cdots, (-1)^N \phi_k)$ through

$$e^{-i\mathbf{p}^k \cdot \mathbf{s}} = \exp\left(i\left(n - \sum_{m=2}^N z_m s_m\right)\phi_k - i\sum_{n=2}^N (-1)^m \phi_k s_m\right) \\ = \exp\left(i\phi_k\left(n - \sum_{m=2}^N (z_m + (-1)^m)s_m\right)\right), \tag{S19}$$

with the integer vector

$$\mathbf{s} = \left(n - \sum_{m=2}^N s_m z_m, s_2, \cdots, s_N\right). \tag{S20}$$

From Eq. (S19), the isotropy of the nearest-neighbor tunneling rates $g^0_{\mathbf{a}_k}$ directly follows. For $n = 0$, Eq. (S19) reduces to

$$e^{-i\mathbf{p}^k \cdot \mathbf{s}} = \exp\left(-i3\phi_k \sum_{m=2}^N \frac{(z_m + (-1)^m)}{3} s_m\right). \tag{S21}$$

Because of our choice of $\omega_m$ in Eq. (14), the quantity $(z_m + (-1)^m)$ appearing in Eq. (S19) is always a multiple of 3 and thus $(z_m + (-1)^m)/3$ is an integer number. Consequently, since the phases $\phi_k$ satisfy

$$e^{-i3\phi_1 q} = e^{-i3\phi_2 q} = e^{-i3\phi_3 q} \tag{S22}$$

for arbitrary integers $q$, we obtain that $\mathcal{J}_0(\mathbf{c}; \mathbf{d}^k)$ and hence $g^0_{\mathbf{a}_k}$ are independent of the direction $k$.

Finally, we will demonstrate the isotropy of the next-nearest-neighbor tunneling rates $\tau_k$ by demonstrating that

$$e^{-i\mathbf{p}^1 \cdot \mathbf{s}} e^{-i\mathbf{p}^2 \cdot \mathbf{s}'} = e^{-i\mathbf{p}^2 \cdot \mathbf{s}} e^{-i\mathbf{p}^3 \cdot \mathbf{s}'} \\ = e^{-i\mathbf{p}^3 \cdot \mathbf{s}} e^{-i\mathbf{p}^1 \cdot \mathbf{s}'}, \tag{S23}$$

with the integer vectors $\mathbf{s}$ in Eq. (S20) and

$$\mathbf{s}' = \left(-n - \sum_{m=2}^{N} s'_m z_m, s'_2, \cdots, s'_N\right), \quad (S24)$$

which implies

$$\mathcal{J}_n^{\mathbf{z}}(\mathbf{c}; \mathbf{d}^1) \mathcal{J}_{-n}^{\mathbf{z}}(\mathbf{c}; \mathbf{d}^2) = \mathcal{J}_n^{\mathbf{z}}(\mathbf{c}; \mathbf{d}^2) \mathcal{J}_{-n}^{\mathbf{z}}(\mathbf{c}; \mathbf{d}^3) = \mathcal{J}_n(\mathbf{c}; \mathbf{d}^3) \mathcal{J}_{-n}(\mathbf{c}; \mathbf{d}^1), \quad (S25)$$

and hence $\tau_1 = \tau_2 = \tau_3$.

In order to prove Eq. (S23), expressions for $e^{-i\mathbf{p}^i \cdot \mathbf{s}} e^{-i\mathbf{p}^j \cdot \mathbf{s}'}$ are necessary. They can be obtained with Eq. (S19) and read

$$e^{-i\mathbf{p}^i \cdot \mathbf{s}} e^{-i\mathbf{p}^j \cdot \mathbf{s}'} = \exp\left(-i3\phi_i \sum_{m=2}^{N} \frac{(z_m + (-1)^m)}{3} s_m\right) \exp\left(-i3\phi_j \sum_{m=2}^{N} \frac{(z_m + (-1)^m)}{3} s'_m\right) e^{in(\phi_i - \phi_j)}. \quad (S26)$$

Eq. (S23) is then derived from Eq. (S26) by using Eq. (S22) and the relation $e^{in(\phi_1 - \phi_2)} = e^{in(\phi_2 - \phi_3)} = e^{in(\phi_3 - \phi_1)}$.

The same result can by readily derived for $\mathbf{F}_-(t)$ by noting that $\mathbf{F}_-(t)$ is obtained from $\mathbf{F}_+(t)$ with the substitution $\omega \to -\omega$ and $\delta_n \to -\delta_n$.

### III. OPTIMAL RESULTS WITH THREE FOURIER COMPONENTS

In the main Letter, we have shown how a suitably chosen driving force $\mathbf{F}_\pm(t)$ with two Fourier components leads to an optimized finite ratio $\frac{j_2}{j_1} \frac{\omega_2}{\omega_0}$ for any desired phase $\phi$. Here, we demonstrate an enhancement of such ratio by increasing the number of frequencies to $N = 3$ and properly choosing the extra driving parameters.

A three-frequency driving force contains a set of driving parameters $\mathbf{p} = (A_1/\omega, A_2/\omega, A_3/\omega, \delta_2, \delta_3)$, where $A_3$ and $\delta_3$ correspond to the amplitude and phase associated to the frequency component $\omega_3 = 4\omega$. Analogously to the two-frequency results, we obtain the optimal choice of $\mathbf{p}$ with the constrained maximization in Eq. (13) [3], with the target functional and constraints expressed analytically in terms of 3-dimensional Bessel functions. In Fig. 1, we display optimal results obtained and compare them with the $N = 2$ results. We observe that, for the entire range of target phases $\phi_{tg}$ except for $\phi_{tg} = \pm\pi/2$, optimally chosen forces with $N = 3$ lead to larger values of $R(\phi_{tg}, r_{th})$ than for bichromatic driving. Moreover, similarly to the $N = 2$ case, we find that the optimal driving force targeting $\phi_{tg} = \pm\pi/2$ corresponds to the monochromatic force, i.e. $A_2 = A_3 = 0$.

Remarkably, the enhancement of the effective next-nearest-neighbor tunneling as compared to the nearest-neighbor tunneling, that is, the enhancement of effects of order $\sim \omega^{-1}$ as compared to effects of order $\sim 1$, relies on the use of a higher frequency component. This strongly suggests that the use of driving forces $\mathbf{F}_\pm(t)$ with even higher frequency harmonics, that is with $N > 3$, and optimally chosen driving parameters could still further improve on the value of $R(\phi_{tg}, r_{th})$. Nonetheless, our results show that already a very low number of frequency components is sufficient to significantly outperform the monochromatic driving and yield significant next-nearest-neighbor tunneling rates for any phase $\phi$.

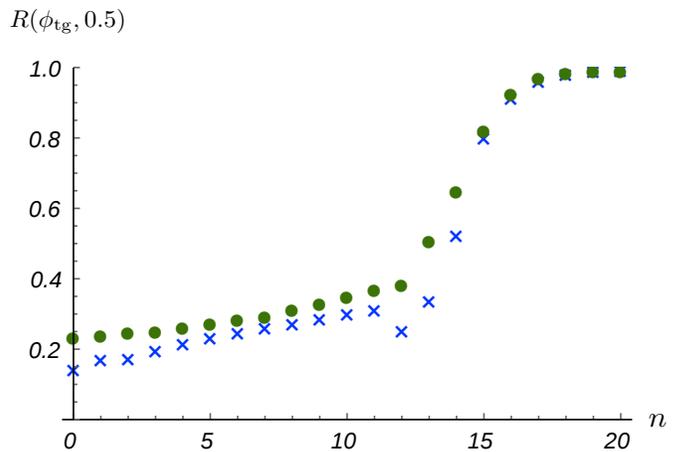

FIG. 1. Plot of $R(\phi_{tg}, r_{th} = 0.5)$ calculated numerically as a function of a discrete set of target phases $\phi_{tg} = \phi_n = n\pi/40$ indexed by $n = 1, \cdots, 20$. Thus, the label $n = 0$ and $n = 20$ correspond to $\phi_{tg} = 0$ and $\phi_{tg} = \pi/2$, respectively. In green, results corresponding to a driving force $\mathbf{F}_+(t)$ with $N = 3$. For comparison, with blue crosses we display again the results for $N = 2$.

demic Press, 1969).

[3] In practice, we do not exactly require constraint $\phi = \phi_{\text{tg}}$ but rather $|\phi - \phi_{\text{tg}}| < \delta\phi$, where $\delta\phi = 0.01$ is the error that we admit.


\end{widetext}

\end{document}